\documentclass[prc,twocolumn,english,showpacs,superscriptaddress]{revtex4}
\usepackage{graphics,epsfig}
\makeatletter

\begin{document}

\title{ Benchmark on neutron capture extracted from $(d,p)$ reactions}
\author{A.M. Mukhamedzhanov}\email{akram@comp.tamu.edu}
\affiliation{Cyclotron Institute, Texas A\& M University,
College Station, TX 77843, USA}
\author{F.M. Nunes}
\affiliation{National Superconducting Cyclotron Laboratory and Department of Physics and Astronomy,
Michigan State University, MI 48824, U.S.A.}
\author{P. Mohr}
\affiliation{Strahlentherapie, Diakonie-Klinikum Schw\"abisch Hall, D-74523
Schw\"abisch Hall, Germany}
\begin{abstract}
Direct neutron capture reactions play an important role in nuclear
astrophysics and applied physics. Since for most unstable
short-lived nuclei it is not possible to measure the $(n, \gamma)$
cross sections, $(d,p)$ reactions have been used as an alternative
indirect tool. We analyze
simultaneously $^{48}{\rm Ca}(d,p)^{49}{\rm Ca}$ at deuteron
energies $2,\,13,\,19$ and $56$ MeV and the 
thermal $(n,\gamma)$ reaction at $25$ meV. We include results for
the ground state and the first excited state of $^{49}$Ca. From
the low-energy $(d,p)$ reaction, the neutron asymptotic normalization
coefficient (ANC) is determined. Using this ANC, we extract
the spectroscopic factor (SF) from the higher energy $(d,p)$ data
and the $(n, \gamma)$ data. The SF obtained through the 
56 MeV $(d,p)$ data are less accurate but consistent with 
those from the thermal capture. 
We show that to have a similar dependence on the single particle
parameters as in the $(n, \gamma)$, the (d,p) reaction
should be measured at $30$ MeV.
\end{abstract}
\maketitle

\section{Motivation}
Reaction rates of capture reactions are a crucial input to
astrophysical network calculations. In particular, neutron capture
reactions, which play a pivotal role in the astrophysical r-process nucleosynthesis,
have to be known for nuclei between the valley of $\beta$-stability and the neutron-drip
line (see \cite{rprocess1,rprocess2,rprocess3} and references therein).
Typically the neutron capture rates for the r-process have been estimated
using the statistical Hauser-Feshbach model, although this may be unreliable away from stability as the level density is low.
$(n,\gamma)$ cross sections for short-lived unstable nuclei cannot
be measured experimentally and have to be taken from theory.
However, the production of unstable nuclei close to the r-process
path has become possible in the recent years, and neutron transfer
experiments like $(d,p)$ on these nuclei are becoming more and
more feasible as beam intensity continues to rise. Experimental
programs using such transfer reactions to derive $(n,\gamma)$
rates (i.e. \cite{jones04,thomas05,thomas07}) have shown promising
results. Theoretical work is needed to place this indirect method
on firmer grounds.
Capture reactions can occur
directly or through a resonance. This work focuses on the direct
capture only and analyzes for the first time $(d,p)$ and $(n,\gamma)$
reactions simultaneously using the combination of spectroscopic
factors (SF) and asymptotic normalization coefficients (ANC)
\cite{muk2005,pang2006,goncharov}. In contrast to charged
particle capture which is mostly peripheral (e.g. \cite{xu,bert02}),
neutron capture reactions may contain an important contribution
from the nuclear interior and consequently be very sensitive to the
spectroscopic factor of the final state \cite{muk2005,pang2006,goncharov}.
One nucleon SFs were first introduced into nuclear physics in the
context of the shell model, where a sequence of orbitals is
generated by a mean-field. SFs provide a measure of the occupancy
of these orbitals \cite{brown88}. Even with modern residual
interactions, shell model predictions appear to overestimate SFs
for well bound systems when compared to experimental values
\cite{brown02,lee06}. The cause for this disagreement is still not
well understood and work along these lines is ongoing. SFs on
unstable nuclei are usually extracted through reactions such as
transfer or knockout \cite{jpg-rev}. Transfer reactions have been
traditionally the prime method of spectroscopy in nuclear physics
(see e.g.\ the recent compilation of $(d,p)$ reactions
\cite{lee07}). In the standard analysis, the experimental cross
section is compared with the predictions from the distorted wave
Born approximation (DWBA) \cite{austern70} or the adiabatic
distorted wave approximation (ADWA) \cite{johnson-ria},
and the SF is extracted from the normalization.
At present, there is a large interest in $(d,p)$ reactions, be it as
an indirect measure of $(n,\gamma)$ rates, as a testing ground for
many-body structure models, or directly associated with stockpile
issues \cite{jolie}. In the conventional approach, 
the extracted SFs suffer from large uncertainty
due to the ambiguity in the bound state and optical potential
parameters. As example we note the recently measured $(d,p)$ reactions
on ${}^{82}{\rm Ge}$ and ${}^{84}{\rm Se}$ nuclei in inverse
kinematics, which have been used to determine the neutron SFs of
$^{83}$Ge and $^{85}$Se above the closed neutron shell at $N = 50$. These
SFs were used to calculate the direct $(n, \gamma)$ capture cross sections
on ${}^{82}{\rm Ge}$ and ${}^{84}{\rm Se}$ \cite{thomas}.  However, the
measured low-energy $(d,p)$ reactions are peripheral; this leads to
large uncertainties in the extracted SFs due to the ambiguity of the
bound state potential parameters, which can deviate significantly from
the standard ones for neutron-rich isotopes.
For all these reasons, the methodology used when
analysing $(d,p)$ data has been recently revisited and questioned
\cite{muk2005,pang2006}. As DWBA predictions are often the base
for the extraction of the phenomenological SFs, sources of
uncertainty in DWBA calculations need to be under control. Once
one has ensured the validity of the one-step approximation
\cite{delaunay}, one needs to assess the uncertainty in the
optical potentials \cite{liu} and the bound state potential
parameters \cite{pang2006}. Along these lines, a combined method
of extracting SFs from transfer reactions, using the asymptotic
normalization coefficients (ANCs) determined independently, was
suggested and tested for different nuclei
\cite{goncharov,pang2006,muk2005}.  Introducing the ANC of the bound
state in the formulation, one controls the normalization of the peripheral part of
the reaction amplitude, generally a large contribution to the transfer
cross section. This allows to significantly reduce the uncertainty
in the choice of the bound state potential parameters and test the
choice of the optical potential or the assumptions in DWBA. These same
ideas can also be applied to direct capture reactions, breakup, and
$\,(e, e'p)$. 
The ultimate goal of this work is to prove the principle that
$(d,p)$ reactions can indeed be used to extract $(n,\gamma)$ rate,
through a systematic methodology. Along these lines, an important
step to validate the use of $(d,p)$ reactions as an indirect tool
to extract $(n, \gamma)$ cross sections for exotic nuclei is to
check for consistency between SFs extracted from $(d,p)$ reactions
and those from $(n, \gamma)$ on stable nuclei, when there is a
large variety of data to constrain the problem. We therefore
benchmark the $(d,p)$ method on $^{48}$Ca$(n,\gamma)^{49}$Ca
(ground state and first excited state). This choice is based on
the very high quality data for thermal neutron capture, the well
known neutron scattering length, and the number of $(d,p)$ data sets
on $^{48}$Ca which all together present a stringent test for the
applied reaction model.
The paper is organized as follows. In section II, we provide some
theoretical background and present the procedure used for
extracting SFs from either $(d,p)$ or neutron capture. In section
III we include the details of the calculations and the data used,
and present the results of our test case. Finally, in section IV
we draw our conclusions.
\section{Theoretical considerations and methodology}
In both $A(d,p)B$ and $A(n,\gamma)B$,  the main nuclear structure
input is the overlap function between the final state and the
initial state $I_{An}^{B}(r)$, its norm being the SF. This
many-body overlap function is usually approximated by a single
particle state $\varphi_{An}(r)$ such that the many-body effects
are hidden into the normalization factor:
\begin{equation}
I_{An}^{B}(r)= S^{1/2}\,\varphi_{An}(r). \label{overlap}
\end{equation}
Here, $S$ is the neutron SF.  We emphasize that Eq.(\ref{overlap})
is an approximation, since the radial dependence of the overlap
function and the nucleon bound state wave function can differ: the
overlap function, as a many body object, includes, in addition to
mean-field effects, the effects of residual interactions, which
affect the nuclear surface region, the relevant region for direct
reactions \cite{mcallen1971}. The single-particle function
$\varphi_{An}(r)$ is the solution of the Schr\"odinger equation
with a central potential, typically a Woods-Saxon potential of
{\it standard} geometry. Asymptotically, the many-body overlap
function and the single-particle function do have the same radial
behaviour. Then, defining $C$ as the asymptotic normalization
coefficient of the overlap function, and $b$ the single-particle
asymptotic normalization coefficient, $C^2=S b^2$. It is clear
that, if one knew $b$ then
knowing the ANC $C$ would provide directly the SF $S$. However, this
is hardly ever the case.  Ambiguities in the single-particle
parameters introduce ambiguities in the SF which are not well controlled
\cite{muk2005,pang2006,xu}.
The DWBA amplitude for the transfer reaction $A(d,p)B$ is given by
$M^{dp}= <\psi_{f}^{(-)}\varphi_{An}|\Delta\,V|\varphi_{pn}\,\psi_{i}^{(+)}>$. The transition operator  is
written in post-form $\Delta V= V_{pn} + V_{pA} - U_{pB}$ with
$V_{ij}$  the interaction potential between $i$ and $j$ and
$U_{pB}$ the optical  potential in the final-state. The distorted
waves in the initial and final states are $\psi^{(+)}_{i}$ and
$\psi^{(-)}_{f}$, and $\varphi_{pn}$ is the deuteron wavefunction.
Similarly, the reaction amplitude for the $(n, \gamma)$ process 
in first order is given by $M^{n\gamma}= <\varphi_{An}|{\hat
O}|\varphi_{scatt}^{(+)}>$, where ${\hat O}$ is the well known
electromagnetic transition operator, and $\varphi_{scatt}$ is 
the neutron incoming scattering wave,  usually calculated with 
the same potential that generates the final neutron bound state
$\varphi_{An}$.  The amplitude for $(n,\gamma)$ is given in the 
usual first order approximation and the operator is taken in the
long-wavelength limit. The $A(d,p)B$ DWBA amplitude depends on the
optical potentials in the initial and final states and the bound
state potential, while the $A(n, \gamma)B$ reaction depends only
on the $V_{An}$ potential for the bound and scattering states.  In
either case, a phenomenological SF is typically extracted through
normalizing the cross section to the corresponding data.
As in \cite{muk2005,pang2006}, and for illustration purposes only, one can
split the total amplitude into an asymptotic part $M_{ext}= b {\tilde
M}_{ext}$, corresponding to $r>R_N$ where the single-particle function
already behaves as a Hankel function, and the remaining interior part
${M}_{int}$.  The cross section is schematically given by
\begin{equation}
\sigma^{\alpha} \propto  S\,|M^{\alpha}[b]|^{2}= |S^{1/2}\,{M}_{int}^{\alpha}[b]  +
C\,{\tilde M}_{ext}^{\alpha}|^{2},
\label{sigma}
\end{equation}
where $\alpha$ stands for $(d,p)$ or $(n, \gamma)$.  Here we express
the dependence on the single particle parameters $(r_0,a)$ by the
single-particle ANC $b=b(r_{0},a)$.  Eq. (\ref{sigma}) brings out
important intuitive physics: the normalization of the external part of
the reaction amplitude is governed by the ANC while the internal part
is determined by the SF. If a reaction is completely peripheral, it is
possible to extract the ANC from the normalization to the data without
any single-particle ambiguity (i.e., no dependence on $b$). In general,
there are contributions from both the internal and external region and
it becomes important to fix the external part independently for an
accurate determination of the SF.  This is true for transfer and
capture reactions.
It is generally said that transfer reactions are surface peaked,
however the internal/external relative contributions can change
significantly with energy. For sub-Coulomb transfer reactions, one is
only sensitive to the asymptotic part of the neutron wavefunction,
and the ANC can be extracted virtually without theoretical uncertainties. 
In \cite{muk2005,pang2006}, the important realization was that,
introducing this independent ANC into the formulation, one could then
extract a SF from a transfer reaction at higher energy, reducing
significantly the uncertainty from the single-particle parameters.
Direct radiative capture $(n, \gamma)$ can be mostly peripheral if
there is a centrifugal barrier in the initial state (e.g.
$^{14}$C$(n,\gamma)^{15}$C \cite{timofeyuk}).  However if there is no
barrier at all ($s-$wave neutron capture), the interior contribution
is important. At typical stellar energies of 10-300 keV, $s-$wave and
$p-$wave capture cross sections are often comparable, see e.g.\
\cite{Mohr26Mg}, so both the ANC and the SF are important ingredients
to calculate the $(n,\gamma)$ cross section.
In this work we will analyse, en par, $(d,p)$ reactions at several
beam energies and the corresponding capture reaction on $^{48}$Ca. By
varying the single-particle radius within a reasonable interval and
fixing the diffuseness, we generate a set of single-particle functions
for a range of single-particle ANCs. For each of these, we calculate
the $(d,p)$ and $(n,\gamma)$ cross sections. Normalizing the peak of
the $(d,p)$ angular distribution to the data at each energy, and
normalizing the thermal neutron capture total cross section, we
calculate phenomenological SFs. We choose $(d,p)$ at sub-Coulomb
energies, where the reaction is known to be peripheral, and normalize
the theoretical cross section to the backward peak in the data. We
include $(d,p)$ at higher energies and normalize the theory to the
first forward peak in the data. In this way we obtain a function
$S_{exp}^{dp}(b)$ from each deuteron energy considered. We also
extract $S_{exp}^{n\gamma}(b)$ from the thermal capture. These
$S_{exp}(b)$ functions obviously vary differently with the
single-particle ANC $b$. If there is consistency in the formulations,
they should all intersect for a realistic $b_0$. This procedure is
better illustrated by looking at the ANCs 
\begin{equation}
[C_{exp}^{dp}(b)]^2 = S_{exp}^{dp}(b) \, b^2 \,\,\, {\rm{and}} \,\,\, 
[C_{exp}^{n\gamma}(b)]^2=S_{exp}^{n\gamma}(b) \, b^2.
\label{eq:SF_ANC}
\end{equation}
If the reaction
is completely peripheral, $C_{exp}(b)$ is constant. If there is
contribution from the interior, then $C_{exp}(b)$ has a slope. This
slope will be more pronounced the larger the interior contribution.
\begin{figure}[tbp]
\epsfig{file=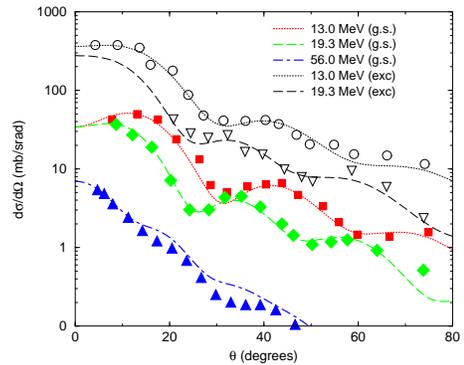,width=6cm}
\caption{
(Color online) Angular distributions for 
${}^{48}{\rm Ca}(d,p){}^{49}{\rm Ca}$ compared to data populating the ground state:
 $E_{d}=13$ MeV (red dotted line), $E_{d}=19$ MeV (green dashed line) 
 and $E_{d}=56$ MeV (blue dot-dashed line). Also shown are the angular distributions to the first excited state:
 $E_{d}=13$ MeV (black thin dotted line) and $E_{d}=19$ MeV (black thin dashed line).
}
\label{fig_dpang} 
\end{figure}
Our test case is $^{49}{\rm Ca}$, which is part of the r-process
chart \cite{rprocess1} and believed to be composed of the double
magic $^{48}$Ca core and a neutron single particle state, either
$p_{3/2}$ for the $^{49}$Ca ground state, or $p_{1/2}$ for the first
excited state in $^{49}$Ca. Direct capture is dominant for
$^{48}$Ca$(n,\gamma)^{49}$Ca because of the small reaction Q-value
and the low level density (see Fig. 4 of \cite{beer1996}). 
These properties of the stable neutron-rich
$^{48}$Ca are close to the expected properties of nuclei close to the
r-process path. For $^{48}$Ca, direct neutron capture occurs via $s-$wave
\cite{beer1996,cranston1971,mohr97,kaepp85} and has a significant
contribution from the nuclear interior. Therefore it is an ideal
test case since the result should be very sensitive to the ANC,
but also to the SF. Another advantage of this choice is that the
scattering length for ${}^{48}{\rm Ca}-n$ has been measured with
sufficient accuracy $a= 0.356 \pm 0.0 88$ fm \cite{sears92} which
allows to determine the initial scattering potential $V_{An}$ with
very minor uncertainties \cite{beer1996}. Neutron radiative
capture on $^{48}{\rm Ca}$
\cite{beer1996,cranston1971,mohr97,kaepp85} and transfer
$^{48}{\rm Ca}(d,p){}^{49}{\rm Ca}$
\cite{rapaport72,metz75,uozumi94} have been accurately measured at
several beam energies, for both the ground state and the first
excited state. Thermal $(n,\gamma)$ data at 25 meV
\cite{cranston1971,beer1996} were measured with $6$\% accuracy,
and sub-Coulomb $(d,p)$ at 1.992 MeV \cite{rapaport72} was
measured with $10$\% accuracy. In this work we also use $(d,p)$
data at 13 MeV and 19.3 MeV from \cite{metz75} and at 56 MeV from \cite{uozumi94}.
\section{Results}
Neutron capture E1 cross sections are determined for the set of bound
state wavefunctions corresponding to the single-particle radius
within the range $r_0=1.05-1.65$ fm and fixed diffuseness
$a=0.65$ fm, with the Woods-Saxon depth $V_{ws}$ adjusted to reproduce
the correct binding energy of the final states in $^{49}$Ca. The depth
$V_{ws}$ for the initial distorted wave is fixed to reproduce the $n
-{}^{48}{\rm Ca}$ scattering length \cite{sears92}.
For the transfer calculations, following the procedure in
\cite{pang2006}, we use the Perey-Perey global optical potential
\cite{perey} to define the optical potential in the exit channel and
construct the deuteron potential in the entrance channel using ADWA
\cite{johnson-ria}, for all but the sub-Coulomb reaction (for the 1.99
MeV $(d,p)$ calculations, we use the optical potentials presented in
\cite{rapaport72} although results are not sensitive to this choice --
e.g.\ a variation of the potential depth by 10\,\% changes the
reaction cross section by only about 1\,\%). ADWA includes deuteron breakup
and thus goes well beyond one-step DWBA. The $n-^{48}$Ca and
$p-^{48}$Ca potentials, needed to construct the finite range adiabatic
deuteron potential \cite{wales}, are also taken from
\cite{perey}. ADWA is developed for reactions where the remnant
$V_{pA} - U_{pB}$ can be neglected. We have checked that this is the
case for $^{48}$Ca$(d,p)$. For $V_{np}$ we use the Reid-soft-core
\cite{rsc} although a simple Gaussian provides the same results.  For
the deuteron bound state we use \cite{rsc} and for $^{49}$Ca, the same
set of bound state wavefunctions used for the $(n,\gamma)$
calculations. All calculations are performed using FRESCO
\cite{fresco}. The shape of the angular distribution at 56 MeV \cite{uozumi94},
in particular in the forward angle region where it is most
important, was not reproduced (this was also the finding in \cite{lee07}). 
For that reason, instead of the Perey and Perey parameterization,
we used the Koning and Delaroche \cite{koning} with which a significant improvement
was obtained for the ground state.
Calculated ADWA angular distributions are compared to
the data in Fig.\ref{fig_dpang}. 
\begin{figure}[tbp] \epsfig{file=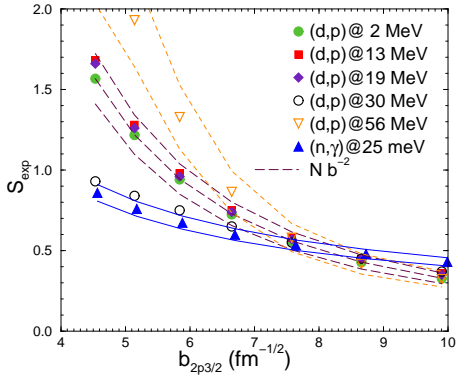,width=6cm}
\caption{(Color online) 
$S_{exp}(b)$ from ${}^{48}{\rm Ca}(d,p){}^{49}{\rm Ca}(g.s.)$ at $E_{d}=1.99$ MeV (green dots),
$E_{d}=13$ MeV (red squares), $E_{d}=19$ MeV (purple diamonds),
$E_{d}=30$ MeV (open circles) and $E_{d}=56$ MeV (open triangles),
and from  ${}^{48}{\rm Ca}(n,
\gamma){}^{49}{\rm Ca}(g.s.)$ at 25 meV (blue triangles). Also
shown are the experimental uncertainties in the $(d,p)$ reaction
at $1.99$ MeV (dashed lines), at $56$ MeV (long-dashed lines) 
and the $(n, \gamma)$ reaction (solid lines). } 
\label{fig_spfgrst1} 
\end{figure} 
\begin{figure}[tbp]
\epsfig{file=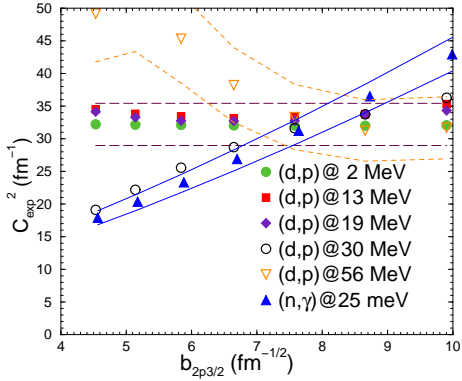,width=6cm} 
\caption{(Color online)
Same as Fig.\ref{fig_spfgrst1} but now referring to the ANC
$C^2_{exp}(b)$. ANC and SF are related by Eq.~(\ref{eq:SF_ANC}).} 
\label{fig_ancgrst1}
\end{figure}
Results for the $^{49}$Ca ground state are shown in
Fig.\ref{fig_spfgrst1} and Fig.\ref{fig_ancgrst1} for
$S_{exp}(b)$ and $C^2_{exp}(b)$ respectively.  As expected, the
sub-Coulomb $(d,p)$ reaction is totally peripheral. At $E_d=1.99$
MeV, the reaction is below the Coulomb barrier ($V_{CB} \approx 5$
MeV) both in the initial and final channels (Q-value is $2.924$
MeV). $C^2_{exp}(b)$ remains constant over the broad interval of
$b$: $C^{2}_{p_{3/2}}= 32.1 \pm 3.2$ fm$^{-1}$ for the
${}^{49}{\rm Ca}(g.s.) \to {}^{48}{\rm Ca} + n$, see
Fig.\ref{fig_ancgrst1}. Consequently the
SF behaves as $1/b^2$ as illustrated in Fig.\ref{fig_spfgrst1}.
The extracted ANC is insensitive to the optical and bound state
potential parameters, the error bar coming from the systematic
error of the data alone. At $E_{d}=13$ and $19$ MeV, the $(d,p)$
reaction turns out to be also dominantly peripheral. It is
reassuring that the $C^{2}_{exp}(b)$ extracted at these energies
\cite{metz75} are consistent with the sub-Coulomb result,
corroborating the ADWA at these energies. However, we should note that, contrary
to the $2$ MeV case, these cross sections are sensitive to the choice
of the optical potentials. The lack of $b$
dependence makes it impossible  to determine the SF from the $13$
and $19$ MeV data. Increasing the beam energy changes the picture.
Fig.\ref{fig_ancgrst1} shows clearly that the 
56 MeV data has a larger contribution from the interior and thus 
is more sensitive to the single particle parameters.
Expectedly, as the single particle ANC increases, so does the external
contribution, which explains the flattening of the $C^2(b)$ curve from 
the $56$ MeV (d,p) data. The error band is represented by
the long-dashed lines in Figs.\ref{fig_spfgrst1} and \ref{fig_ancgrst1}.
The cross sections at this energy were measured
with $10$\% accuracy but we have added in quadrature another $10$\% due 
to the dependence on the optical potentials. From the joint analysis of the 2 MeV data and the 56 MeV data,
we can  extract a spectroscopic factor $S_{exp}= 0.55 \pm 0.25$.
The results for the ANC extracted from ${}^{48}{\rm Ca}(n,
\gamma){}^{49}{\rm Ca}(g.s.)$ show a strong $b$ dependence, in
perfect agreement with \cite{baye04}, confirming an important contribution
from the nuclear interior.  From the overlapping
regions in Fig.\ref{fig_ancgrst1}, we obtain $b_{0}= 7.80 \pm 1.2$ fm$^{-1/2}$
and $S_{exp}= 0.53 \pm 0.11$, in agreement with the value obtained
from the $56$ MeV transfer data. This $b_0=7.80$ fm$^{-1/2}$ 
corresponds to $r_0 \approx 1.45$ fm, a radius larger than the standard value.  
The determined SF is lower than one could expect from the independent
particle shell model, confirming the reduction of the SF previously
seen \cite{lee06}, and now the ambiguity of the bound state
potential parameters has been eliminated. 
One may ask whether there are deuteron energies $20<E_d<56$ (MeV) for which
the $(d,p)$ cross sections show the same sensitivity to the
nuclear interior as the corresponding $(n,\gamma)$. We find that
$E_d=30$ MeV provides an excellent match. ADWA is expected to
perform well in this energy region, as the adiabatic condition is
satisfied (see \cite{liu} for a successful application on
$^{12}$C). We thus consider the $(d,p)$ reaction at $E_d=30$ MeV
and apply an arbitrary normalization to the cross section (open
circles in Fig.\ref{fig_spfgrst1} and Fig.\ref{fig_ancgrst1}).
Clearly illustrated in
Fig.\ref{fig_ancgrst1}, $C^2_{exp}(b)$ obtained from the $(d,p)$
reaction at $30$ MeV has a similar $b$ dependence as the
$C^2_{exp}(b)$ from the $(n, \gamma)$ process.
We emphasize that a $(d,p)$ experiment on $^{48}$Ca at this
energy would be very useful; plans for such an experiment have been
initiated by the present study \cite{Tribb_priv}.
\begin{figure}[tbp]
\epsfig{file=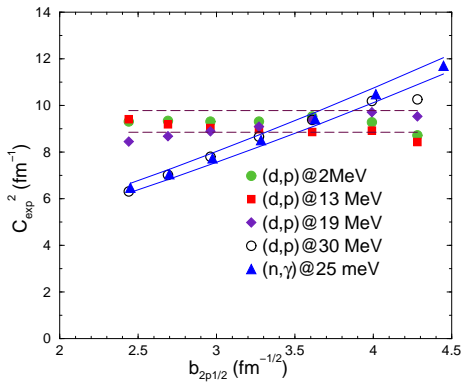,width=6cm} 
\caption{(Color online)
Same as Fig.\ref{fig_ancgrst1} but now referring to first
excited state in $^{49}$Ca.} 
\label{fig_ancexst1}
\end{figure}
We also study the $(d,p)$ and $(n, \gamma)$ reactions to the first
excited state in ${}^{49}{\rm Ca}$. Results for $S_{exp}(b)$ and
$C^2_{exp}(b)$ show the same pattern as for the ground state. Because
of the relation between ANC and SF we show only
$C^2_{exp}(b)$ in Fig.\ref{fig_ancexst1}. We obtain from
sub-Coulomb $(d,p)$ reaction, $C^{2}_{p_{1/2}}= 9.30 \pm 0.93$
fm$^{-1}$. Reactions at 13 and 19 MeV remain peripheral and the 
$56$ MeV distribution is not well described by our model. Again,
the $(d,p)$ at 30 MeV and the $(n, \gamma)$ show a similar sensitivity
to the interior as for the ground state transition.  Our sub-Coulomb
$(d,p)$ and $(n, \gamma)$ joint analysis provides $b_0 = 3.63^{+0.54}_{-0.58}$
fm$^{-1/2}$ and $S_{exp} = 0.71^{+0.20}_{-0.12}$. Again, this
$b_0$ corresponds to $r_0 \approx 1.45$ fm, showing consistency in
the geometry of the neutron $p_{3/2}$ and $p_{1/2}$ orbitals.
The large radii $r_0$ give further evidence for a neutron skin in
neutron-rich calcium isotopes \cite{bert07}.
\section{Conclusions}
Our goal was to provide a proof of principle that $(d,p)$
reactions can indeed be used to extract $(n,\gamma)$ rates, through
a systematic methodology. As sub-Coulomb or near-Coulomb-barrier
$(d,p)$ reactions are peripheral, they provide an ANC virtually
free from optical potential ambiguities. Whenever the $(n,
\gamma)$ reaction is peripheral, this is the only necessary bound
state information needed for calculating the neutron capture cross
section at low energy. If the $(n, \gamma)$ is not peripheral, one
needs to know both, the ANC and the SF. We have shown for $^{48}$Ca
that SFs obtained from thermal capture and $56$ MeV (d,p)
are consistent. Our results suggest that $(d,p)$ data at energies around $E_d=30$
MeV has a similar interior contribution as the $(n,\gamma)$ and could
thus be a better tool for extracting the $(n,\gamma)$ cross sections. 
In order to confirm this, experiments at 30 MeV providing cross sections with better
than $10$\% accuracy, would be very useful. 
Any prediction of neutron capture cross sections on unstable nuclei 
will be complicated by the fact that the neutron scattering length will 
not be measurable in most cases. This leads to an additional uncertainty 
because the scattering potential is not as well-defined as for the 
presented example $^{48}{\rm Ca}-n$. A rough estimate (see e.g.\ Fig.~5 of 
Ref.~\cite{beer1996}) shows that this additional uncertainty remains 
below a factor of two for realistic variations of the neutron scattering 
potential. A detailed study of uncertainties of neutron capture cross 
sections for unstable nuclei will be given in a forthcoming paper.

From this work, we also find that $s-$wave $(n, \gamma)$ reactions
are actually very well suited for extracting a SF. For the $(n,
\gamma)$ process, the transition operator is well known, the
ambiguity in the optical potential at very low energies is
negligible when imposing the correct neutron scattering length,
and the ambiguity of the bound state potential parameters is
greatly reduced by fixing the asymptotic part of the bound state,
say through sub-Coulomb reactions.
Hence, the joint analysis of low energy $(d,p)$ and thermal
$s-$wave $(n,\gamma)$ provides a powerful method to test
microscopic structure models.
{\small This work is supported by the NNSA-DOE Grants
DE-FG02-93ER40773 and DE-FG52-03NA00143 with a Rutgers subcontract DE-FG52-06NA26207, 
and the National Science Foundation, under grant PHY-0555893.}

\end{document}